\documentclass[aps,pre,10pt,superscriptaddress,nofootinbib,letterpaper,twocolumn]{revtex4-2}
\usepackage{amsfonts,amssymb,amsmath,bm,bbm,mathrsfs}
\usepackage[mathcal]{euscript}
\usepackage[usenames]{color}
\usepackage{color,soul}
\usepackage[utf8]{inputenc}
\usepackage[T1]{fontenc}
\usepackage{textcomp}
\usepackage{graphicx}
\usepackage[hypertex]{hyperref}
\newcommand{\rmd}{\mathrm{d}}

\newcommand{\mrmp}{\mathrm{p}}

\newcommand{\bfx}{\mathbf{x}}
\newcommand{\bfe}{\mathbf{e}}

\newcommand{\bfz}{\mathbf{z}}

\begin{document}

\title{Small equatorial deformation of homogeneous spherical fluid vesicles}

\author{Andrés Solís-Cuevas}
\affiliation{Facultad de Matemáticas, Universidad Autónoma de Yucatán,\\
Periférico Norte, Tablaje 13615, CP 97110, Mérida, Yucatán, México}
\author{Pablo Vázquez-Montejo}
\email[]{pablo.vazquez@secihti.mx}
\email[]{pablo.vazquez@correo.uady.mx}
\affiliation{SECIHTI - Facultad de Matemáticas, Universidad Autónoma de Yucatán,\\
Periférico Norte, Tablaje 13615, CP 97110, Mérida, Yucatán, México}

\begin{abstract}
We examine the reaction of a homogeneous spherical fluid vesicle to the force exerted by a rigid circular ring located at its equator in the linear regime. We solve analytically the linearized first integral of the Euler-Lagrange equation subject to the global constraints of fixed area and volume, as well as to the local constraint imposed by the ring. We determine the first-order perturbations to the generating curve of the spherical membrane, which are characterized by the difference of the radii of the membrane and the ring, and by a parameter depending on the physical quantities of the membrane. We determine the total force that is required to begin the deformation of the membrane, which gives rise to a discontinuity in the curvature of the membrane across the ring.
\end{abstract}

\maketitle

\section{Introduction}

Fluid membranes display a broad class of configurations with different geometric and topological features. They can be deformed by changing the physical parameters of their environment, such as the pressure or temperature, or due to their interaction with proteins or polymers \cite{Seifert1997, LipowskyBook, ThalmannBook}, physical situation that is relevant in many processes of cellular division \cite{Reynwar2007, Lenz2009, Frolov2015, Renard2018}. One cellular fission process that has been widely studied is the cytokinesis, in which actin filaments polymerize on the plasma membrane, assembling into an equatorial contractile ring which exerts an inward force, forming a cleavage furrow that facilitates the symmetric fission of the membrane into two vesicles \cite{Cortes2018, Pollard2019}. The mechanics of this process has been studied using different approaches: hydrodynamic theory of active gels \cite{Julicher2007, Joanny2009}, perturbatively using a trigonometric series ansatz  \cite{Almendro2013, Beltran2017, Beltran2019}, as well as numerical methods, simulations and triangulations \cite{Biron2005, Cumberworth2023, Okuda2023}. Also, the constriction of fluid vesicles by a ring of Janus nanoparticles has been examined using Monte Carlo simulations \cite{Bahrami2019}.
\\
Another example of membranes subject to external forces that have been studied theoretically, numerically and experimentally are tethered membranes. By analyzing the response of fluid vesicles to point forces exerted at their poles it has been found that they adopt a variety of  axisymmetric configurations: pulling their poles the vesicles elongate, achieving cylindrical and spindle shapes, whereas if their poles are pushed they attain dimpled shapes \cite{Bozic1997, Heinrich1999, Powers2002, Leroy2021, Reboucas2024}. The latter case has also been studied theoretically by exploiting the conformal invariance of the bending energy: the inversion in spheres of the catenoid results in vesicles with two points in contact, where the compressive forces originate logarithmic singularities in the membrane curvature \cite{Castro2007A, Castro2007B, Guven2013}.
\\
In the variational approach, the equilibrium configurations of fluid membranes are described by the solutions of the Euler-Lagrange (EL) equation that is obtained by minimizing their bending energy. Since it is a fourth order differential equation in the embedding functions of the membrane, the study of its analytic solutions narrows down to simplified situations, for instance, configurations with axial symmetry (which entails the existence of a first integral of the EL equation), such as spheres, cylinders and the Clifford torus; or configurations with constant mean curvature, such as minimal surfaces or Delaunay surfaces, \cite{Naito1995, Seifert1997, Osserman1986, FomenkoBook, Kenmotsu2003, LipowskyBook}. Furthermore, next to nothing is known about analytical solutions which take into account the interactions of membranes with filaments, so in order to consider more general or realistic situation one has to resort to simulations or numerical methods. For instance, the first integral of the EL equation has been solved numerically to study the equatorial deformation of homogeneous fluid vesicles with fixed area or fixed volume by a rigid ring in the nonlinear regime \cite{Vazquez2025}. However, more often than not, perturbation theory can be employed to gain insight into the initial behavior or instabilities of the physical system under consideration, either for elastic rods or membranes. The prototypical example is the well-known Euler buckling instability of a straight rod, which bends under a compressive force whose magnitude is greater than a critical value \cite{LandauBook, AudolyBook}. Likewise, straight, circular and helical Kirchhoff rods buckle if they are twisted beyond a critical value \cite{Goriely2000, Solis2021}. Above a threshold pressure difference a spherical vesicle exhibits small deformation instabilities, which are characterized by an infinite number of modes of deformation with a discrete spectrum \cite{ZhongCan1987, ZhongCan1989, Seifert1991}. Also, the geometry of narrow necks of homogeneous fluid vesicles, along with the force constricting them, can be determined employing a perturbative analysis about the catenoid \cite{Vazquez2025}.
\\
In this work, we analyze perturbatively the initial stage of the equatorial deformation of a spherical fluid membrane by a circular rigid ring. To this end we employ the spontaneous curvature model adapted to axial symmetry. In Sect. \ref{sectaxipar} we define the nondimensional parameters and geometric quantities of the axisymmetric parametrization that we use to describe the membrane configurations. In Sect. \ref{sectHEL} we review the first integral of the EL equation, which stems from the rotational symmetry of the bending energy of the vesicle, along with the appropriate boundary conditions that are used to determine the equilibrium configurations in the nonlinear regime. In Sect. \ref{Sectspherepertdef} we develop the perturbative analysis about the sphere. We derive the linearized system of equations whose solutions represent the first order change in the geometry of the spherical membrane under a small change of its equatorial radius. The small parameter of the deformation is given by the difference of the radii of the ring and the initial spherical vesicle: if it is positive (negative)  the vesicle is stretched (constricted).
We  derive the appropriate boundary conditions and the constraints that the first order corrections should satisfy due to the presence of the ring and to preserve the total area and volume of the vesicle. We examine the first order changes in the total energy of the membrane and the total force exerted by the ring. In Sect. \ref{SectWeakConstSphere} we present the analytic solutions of the linearized system of differential equations, which are characterized by a parameter that depends on the physical quantities of the membrane. We consider three cases, two of them corresponding to constant values of such parameter, and a general case where the parameter is different from those two constant values. We determine the first order corrections to the radial and height coordinates for these three cases, as well as the critical value of the external force that is required to initiate such small deformation of the spherical vesicle, which is reflected by a discontinuity in the membrane curvature along the  orthogonal direction to the ring. We end with the discussion and conclusions of our analysis in Sect. \ref{SecDisConc}. Some formulas that are useful for solving the linearized system of differential equations are collected in Appendices \ref{AppA} and \ref{AppB}.

\section{Geometric quantities of axisymmetric membranes} \label{sectaxipar}

\begin{figure}[htb]
\centering
 \includegraphics[scale=0.65]{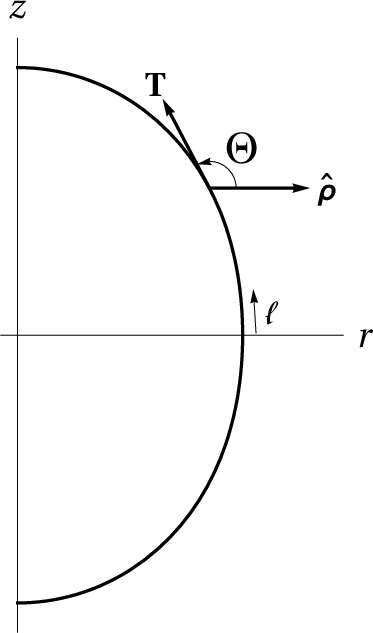}
 \caption{The profile curve of the axisymmetric surface is defined by the scaled radial and height coordinates $r$ and $z$, which are parametrized by the scaled arc length $\ell$. The tangent vector $\mathbf{T}$ forms an angle $\Theta$ with the radial direction $\hat{\bm{\rho}}$.}
 \label{fig:1}
\end{figure}

We consider the deformation of a spherical vesicle of radius $R_S$ and spontaneous curvature $C_s$ by a circular rigid ring of radius $R_0$ placed at the equator, such that the axial and equatorial mirror symmetries are preserved. The axisymmetric surface representing the vesicle is parametrized by the azimuthal angle $\phi$ and the arc length $l$ of the generating curve, whose radial and height coordinates are denoted by $R(l)$ and $Z(l)$.
\\
The radius $R_S$ of the spherical vesicle sets the relevant length scale of the problem. In the following, for convenience we work with nondimensional quantities, so we rescale all the quantities and parameters of the vesicle with appropriate powers of $R_S$. The arc length, the radial and height coordinates are scaled with $1/R_S$, \begin{equation}
\ell:=\frac{l}{R_S}\,, \quad r:=\frac{R}{R_S}\,, \quad z := \frac{Z}{R_S}\,.
\end{equation}
On account of the equatorial mirror symmetry of the membrane, we need to consider only one half of the generating curve, whose scaled arc length is measured from the equator, where $\ell=0$, up to the pole where $\ell=\ell_p$. We use the scaled coordinates parametrized by scaled arc length, $r(\ell)$ and $z(\ell)$, to describe the generating curve of the membrane (Fig. \ref{fig:1}). Furthermore, we employ as an auxiliary variable the angle $\Theta$ that the unit tangent vector $\mathbf{T}$ of the generating curve makes with the radial direction $\bm{\hat{\rho}}$ (shown in Fig. \ref{fig:1}), defined by
\begin{equation} \label{EqsRZ}
\dot{r} = \cos \Theta\,, \quad \dot{z} = \sin \Theta\,,
\end{equation}
where the dot represents differentiation with respect to scaled arc length ($\,\dot{} := d/d\ell$).
\vskip0pc \noindent
The area element $\rmd A$ is scaled by $1/R_S^2$,
\begin{equation}
\rmd a := \frac{\rmd A}{R_S^2} = r \rmd\phi \wedge \rmd\ell \,.
\end{equation}
The total area $A$ and total volume $V$ of the vesicle are scaled with the total area and total volume of the initial spherical vesicle, $A_S = 4 \pi R_S^2$ and $V_S= (4\pi/3) R_S^3$, which for a membrane with symmetry with respect to the equatorial plane are given by
\begin{subequations}
\begin{eqnarray}
a &:=& \frac{A}{A_S} = \int_0^{\ell_p} r \, \rmd \ell \,, \\
v&:=&\frac{V}{V_S} = \frac{3}{2} \int_0^{\ell_p} r^2 \sin \Theta \, \rmd \ell \,.
\end{eqnarray}
\end{subequations}
The principal curvatures $C_\perp$ and $C_\parallel$, corresponding to the curvatures of the meridians and parallels, are scaled by $R_S$,
\begin{subequations}
\begin{eqnarray}
c_\perp &:=& R_S C_\perp = \dot{\Theta} \,, \\
c_\parallel &:=& R_S C_\parallel = \frac{\sin \Theta}{r} \,.
\end{eqnarray}
\end{subequations}
Twice the mean curvature $K$ and the Gaussian curvature $K_G$ are scaled by $R_S$ and $R_S^2$ respectively
\begin{subequations}
\begin{eqnarray}
k&:=& R_S K = c_\perp +c_\parallel \,, \\
k_G&:=&R_S^2 K_G =c_\perp c_\parallel\,.
\end{eqnarray}
\end{subequations}
We define the scaled spontaneous curvature by $ c_s = R_S C_s$, as well as the difference between the scaled mean and spontaneous curvatures,
\begin{equation}  \label{def:kappa}
 \kappa = k - c_s \,.
\end{equation}
It proves useful to regard $\kappa$ also as an auxiliary variable in the determination of the equilibrium configurations of the membrane. From Eq. (\ref{def:kappa}) we obtain a first-order differential equation for $\Theta$,
\begin{equation} \label{EqTheta}
\dot{\Theta} = \kappa -\frac{\sin \Theta }{r} + c_s \,.
\end{equation}
Equations (\ref{EqsRZ}) and (\ref{EqTheta}), are part of the system of differential equations that determines the configuration of the membrane. The missing equation for $\kappa$ is obtained from the variational principle.

\section{Bending energy and first integral of the Euler-Lagrange equation} \label{sectHEL}

We consider the total bending energy of the vesicle, which according to the spontaneous curvature model is quadratic in $\kappa$ \cite{Canham1970, Helfrich1973, Deserno2015}, scaled with $1/H_{B S}$, where $H_{B S}= 8 \pi k_B$ is the total bending energy of a spherical vesicle with null spontaneous curvature and $k_B$ is the bending modulus. The scaled total bending energy of an axisymmetric membrane with equatorial mirror symmetry is given by
\begin{equation} \label{def:Htot}
 h_B = \frac{1}{16 \pi} \int \kappa^2 \, \rmd a = \frac{1}{4} \int_0^{\ell_p} \kappa^2 r \, \rmd\ell \,.
\end{equation}
We do not include a term proportional to the surface integral of the Gaussian curvature, because it is a topological invariant due to the Gauss-Bonnet theorem \cite{FrankelBook, DoCarmoBook}, so it would only contribute with a constant term to the total energy.
\\
The deformation of the spherical vesicle ought to fulfill two physical constraints reflecting the constant number of amphiphilic molecules constituting the membrane, as well as the incompressibility of the enclosed volume. These constraints imply that the total scaled area and total volume of the membrane are fixed, say to values $a_0=A_0/A_S$ and $v_0=V_0/V_S$, respectively. In the following, we consider deformations preserving the area of the initial spherical configuration, so the scaled total area and volume are unit, $a_0=1$ and $v_0=1$. Both constraints are enforced in the variational principle with two global nondimensional Lagrange multipliers, $\mu= R_S^2 \sigma/k_B$ and $\mrmp = R_S^3 P/k_B$, where $\sigma$ and $P$, are the intrinsic surface tension and osmotic pressure difference across the membrane.\footnote{These constraints could be considered separately, for instance, the case of fixed area but variable volume, corresponding to $\mrmp = 0$, could represent the situation of isothermal transformations, where the osmotic pressure would be the control parameter. By contrast, the case of isochoric transformations, i.e. fixed volume but variable area, corresponding to $\mu=0$, could represent the case where the temperature would be the control parameter \cite{Seifert1997, LipowskyBook}.} Also, the presence of the ring at the equator constrains the scaled equatorial radius of the membrane $r(0)=R(0)/R_S$ to be equal to its scaled radius $r_0=R_0/R_S$, condition that is also implemented using a scaled Lagrange multiplier $\phi_0=R_S^2 \lambda /k_B$, where $\lambda$ is the linear force density exerted by the ring. Thus, the constrained energy (scaled with $1/H_{BS}$) to be minimized, is given by
\begin{eqnarray}
 h_C &=& h_B + \frac{\mu}{2} (a-1) - \frac{\mrmp}{6} (v-1) \nonumber \\
 &+& \frac{1}{8 \pi} \oint \phi_0 (r(0)-r_0) \rmd s\,,
\end{eqnarray}
where $d s = r_0 d \phi$ is the line element of the equator scaled with $1/R_S$. The axial symmetry of $h_B$ implies the existence of a first integral of the corresponding EL equation, which is a first-order differential equation for $\kappa$ (details of the derivation can be found in Ref. \cite{Vazquez2025})
\begin{eqnarray} \label{Eqkappa}
\cos \Theta \, \dot{\kappa} &+& \sin \Theta \left( \kappa \left(
\frac{\kappa}{2} -\frac{\sin \Theta}{r}+ c_s \right) - \mu  \right) \nonumber \\
&+& \frac{\mrmp}{2} r = 0\,.
\end{eqnarray}
The equilibrium configurations are determined by solving the system of first order differential equations for $r$, $z$, $\Theta$ and $\kappa$ given by Eqs. (\ref{EqsRZ}), (\ref{EqTheta}) and (\ref{Eqkappa}).
In order to determine completely the solutions of the system of differential equations we need five conditions which allow for the specification of the four constants of integration and one of the Lagrange multipliers (the other one can be obtained from a scaling relation). In the nonlinear regime, the numerical solutions are determined by imposing the following four boundary conditions (BCs)
\begin{subequations}  \label{eq:BCsnonlinreg}
\begin{align}
 r(0) &= r_0 \,, &\quad z(0)&=0\,, \label{eq:BCrz} \\ 
\Theta(0) &= \frac{\pi}{2} \,, &\quad \Theta (\ell_p)& = \pi \,. \label{eq:BCThetaeqpls}
 \end{align}
\end{subequations}
The first two BCs reflect the presence of the ring, whose radius fixes the initial value of $r$ at the equator, which is located on the plane with $z=0$. The other two BCs entail the existence of tangent planes at the equator and the poles, such that the bending energy remains finite. Thus, in principle this is a boundary valued problem, where the BC for $\Theta$ replaces the specification of the initial value of $\kappa$ at the equator, $\kappa(0)$. However, instead of implementing the BC for $\Theta$ at the pole, the shooting method is used, and $\dot{\Theta}_0$ is tuned such that this BC is fulfilled. Furthermore, since the first BC for $\Theta$ at the equator is inconsistent with  Eq. (\ref{Eqkappa}), which is valid only for membranes free of external forces, an approximate initial value is used $\Theta(0) \approx \pi/2$, \cite{Vazquez2025}.
\\
The Lagrange multipliers $\mu$ and $\mrmp$ are determined from the conditions of constant total area or constant total volume of the vesicle. The total scaled equatorial external force $f_0$ is determined from the line integral along the equator of the scaled linear force density $\phi_0$,
\begin{equation}
 f_0 = \oint \phi_0 \rmd s\,.
\end{equation}
The magnitudes of the scaled force density and the total force at the equator for a vesicle with equatorial mirror symmetry are given by \cite{Bozic2014, Guven2018, Vazquez2025}
\begin{equation} \label{Eq:f0}
\phi_0 = -2 \ddot{\Theta}_0\,, \quad
f_0 = -4 \pi r_0 \ddot{\Theta}_0 \,,
\end{equation}
where $\ddot{\Theta}_0:=\ddot{\Theta}(0)$ is the value of the second derivative of the angle
$\Theta$ at the equator. The sign of $f_{0}$ indicates the direction of the total force; $f_{0}<0$ ($f_{0}>0$) corresponds to a constrictive (dilative) force.

\section{Perturbative expansion about the sphere} \label{Sectspherepertdef}

In order to analyze the initial deformation of a spherical vesicle, we employ a perturbative analysis about the sphere. The scaled radial and height coordinates and angle are given by
\begin{subequations} \label{sphericalsol}
\begin{eqnarray}
 r_{(0)}(\ell) &=& \cos \ell \,, \\
 z_{(0)}(\ell)&=&\sin \ell\,, \\
 \Theta_{(0)}(\ell) &=& \ell + \frac{\pi}{2}\,,
\end{eqnarray}
 \end{subequations}
where the scaled arc length is in the interval $0\leq \ell \leq \ell_p =  \pi/2$. The scaled principal curvatures of the sphere are unit
\begin{equation}
c_{\parallel(0)} =  c_{\perp(0)} =1 \,
\end{equation}
Thus, for the sphere, the scaled mean curvature difference is constant
\begin{equation}
\kappa_{(0)} =  2 - c_s \,.
\end{equation}
The sphere is a solution of Eq. (\ref{Eqkappa}) whose parameters satisfy the following
equation \cite{ZhongCan1987, ZhongCan1989, Seifert1991}
\begin{equation} \label{eq:spheresclrl}
\mu_{(0)} - \frac{\mrmp_{(0)}}{2}  + c_s \left( \frac{c_s}{2}-1 \right)=0 \,.
\end{equation}
We write $r$, $z$, $\Theta$ and $\kappa$, using a perturbative expansion
\begin{subequations} \label{eqs:exprzThkappa}
\begin{eqnarray}
r &=& r_{(0)} + r_{(1)} + \dots \,, \\
z &=& z_{(0)} + z_{(1)} + \dots \,, \\
\Theta &=& \Theta_{(0)} + \Theta_{(1)} +\dots\,, \\
\kappa &=& \kappa_{(0)} + \kappa_{(1)} +\dots \,,
\end{eqnarray}
\end{subequations}
where the zeroth order terms are given by Eqs. (\ref{sphericalsol}). Likewise, we expand the parameters as
\begin{subequations} \label{eqs:expmup}
\begin{eqnarray}
\mu &=& \mu_{(0)} + \mu_{(1)} +\dots \,, \\
\mrmp &=& \mrmp_{(0)} + \mrmp_{(1)} +\dots\,,
\end{eqnarray}
\end{subequations}
where the zeroth order terms satisfy Eq. (\ref{eq:spheresclrl}). Since the reduced spontaneous curvature $c_s$ is a given finite parameter, it only enters at zeroth order, thus $\kappa_{(1)}=k_{(1)}$.
\\
The scaled embedding functions are expanded as
\begin{subequations}
\begin{eqnarray}
\bfx &=&\bfx_{(0)} + \bfx_{(1)} + \dots \,, \\
\bfx_{(0)} &=& r_{(0)} \hat{\bm{\rho}} + z_{(0)} \hat{\bfz} \,, \\
\bfx_{(1)} &=& r_{(1)}  \hat{\bm{\rho}} + z_{(1)} \hat{\bfz} \,.
\end{eqnarray}
\end{subequations}
Thus, the tangent vector, $\bfe_\ell = \dot{\bfx}$, is expanded as
\begin{subequations} \label{eq:elser}
 \begin{eqnarray}
\bfe_\ell &=& \bfe_{\ell(0)} + \bfe_{\ell(1)} + \dots\,, \\
\bfe_{\ell(0)} &=& \dot{r}_{(0)} \hat{\bm{\rho}} + \dot{z}_{(0)} \bfz \,, \\
\bfe_{\ell(1)} &=& \dot{r}_{(1)} \hat{\bm{\rho}} + \dot{z}_{(1)} \bfz \,.
 \end{eqnarray}
\end{subequations}
Substituting the perturbation series given in Eqs. (\ref{eqs:exprzThkappa}) and (\ref{eqs:expmup}), we obtain that to first order Eqs. (\ref{EqsRZ}), (\ref{EqTheta}) and (\ref{Eqkappa}) read
\begin{subequations} \label{eq:sphere1stord}
 \begin{eqnarray}
\dot{r}_{(1)} + \cos \ell \, \Theta_{(1)}  &=&0\,, \label{sphedr} \\
\dot{z}_{(1)} + \sin \ell \, \Theta_{(1)} &=&0\,, \label{sphedz} \\
\dot{\Theta}_{(1)} - \sec \ell \, \left( \sin {\ell} \Theta_{(1)} + \, r_{(1)} \right) - \kappa_{(1)} &=&0 \,, \label{sphedTheta} \\
- \sin \ell \, \dot{\kappa}_{(1)} + \cos \ell \left( \kappa_{(1)}  - C_{(1)} \right) && \nonumber \\
+ \left(2-C_{(0)}\right)  \left( r_{(1)} + \sin \ell \Theta_{(1)} \right) &=&0 \,, \label{sphedkappa}
\end{eqnarray}
\end{subequations}
where we have defined the following constants
\begin{subequations}
\begin{eqnarray}
 C_{(0)} &=& c_s \left(2 - \frac{c_s}{2}\right)-\mu_{(0)} = c_s -\frac{\mrmp_{(0)}}{2}
 \,, \label{def:C0}\\
 C_{(1)} &=& \mu_{(1)} - \frac{\mrmp_{(1)}}{2}\,. \label{def:C1}
\end{eqnarray}
\end{subequations}
We have used Eq. (\ref{eq:spheresclrl}) in the second equality of Eq. (\ref{def:C0}). Using Eqs. (\ref{sphedr}) and (\ref{sphedz}) for the derivatives of $r_{(1)}$ and $z_{(1)}$ and Eqs. (\ref{eq:elser}) for the expansion of the tangent vector, we have that $\bfe_{\ell(0)} \cdot \bfe_{\ell(1)}=0$, thus, to first order $\lVert \bfe_\ell \rVert = 1$, so there is no first-order correction to the scaled arc length, and it is still in the interval $0\leq \ell \leq \pi/2$.
\\
Equation (\ref{sphedTheta}) provides the first-order correction to the mean curvature $\kappa_{(1)}$, given by the sum of the first order corrections to the curvatures,
\begin{equation} \label{Eqs:frstordcprpcpar}
c_{\perp(1)} =\dot{\Theta}_{(1)} \,, \quad  c_{\parallel(1)} =-\frac{r_{(1)}+\sin \ell \Theta_{(1)} }{\cos \ell}\,.
\end{equation}
Since the scaled principal curvatures are unit, the first-order correction to the Gaussian curvature is equal to the scaled mean curvature
\begin{equation} \label{Eq:kG1}
k_{G(1)} = c_{\parallel(0)}  c_{\perp(1)} + c_{\perp(0)} c_{\parallel(1)} = \kappa_{(1)} \,.
\end{equation}
In order to render $c_{\parallel(1)}$ finite at the pole with $\ell=\pi/2$, the numerator ought to vanish. There are two possibilities, either the first-order corrections vanish separately, $r_{(1)p}:=r_{(1)}(\pi/2)=0$ and $\Theta_{(1)p}:=\Theta_{(1)}(\pi/2)=0$, or they cancel each other, $r_{(1)p}+\Theta_{(1)p}=0$.
\\
Now we derive a differential equation for $\kappa_{(1)}$ that provides a starting point for solving this system of differential equations. First, combining Eqs. (\ref{sphedTheta}) and (\ref{sphedkappa}) we can eliminate the quantity $r_{(1)}+ \sin \ell \Theta_{(1)}$ and obtain $\dot{\Theta}_{(1)}$ in terms of $\kappa_{(1)}$ and its arc length derivative
\begin{equation} \label{DETheta1p}
 (2-C_{(0)})\dot{\Theta}_{(1)} = \tan \ell \, \dot{\kappa}_{(1)} + (1-C_{(0)}) \kappa_{(1)} + C_{(1)}\,.
\end{equation}
Differentiating Eq. (\ref{sphedkappa}) and using Eq. (\ref{sphedr}) we get another equation for $\Theta_{(1)}$ in terms of the second order derivative of $\kappa_{(1)}$
\begin{equation} \label{DETheta1pddkappa}
(2-C_{(0)}) \dot{\Theta}_{(1)} = \ddot{\kappa}_{(1)} + \kappa_{(1)} - C_{(1)}\,.
\end{equation}
Combining Eqs. (\ref{DETheta1p}) and (\ref{DETheta1pddkappa}) we get a second-order differential equation for $\kappa_{(1)}$
\begin{equation} \label{Eq:ddkappa1}
\ddot{\kappa}_{(1)} - \tan \ell \, \dot{\kappa}_{(1)} + C_{(0)} \kappa_{(1)} - 2 C_{(1)}=0\,.
\end{equation}
In the following, we consider three types of solutions depending on the value of the constant $C_{(0)}$: $C_{(0)}=0,2$ and $C_{(0)}\neq 0,2$.
\\
To solve this system of equations we need five conditions to determine the four constants of integration and one of the Lagrange multipliers. The nonlinear BCs (\ref{eq:BCsnonlinreg}) imply that the first-order corrections to $r$, $z$ and $\Theta$ should satisfy the following BCs at the equator
\begin{subequations} \label{BCsrzTheta}
\begin{eqnarray}
 r_{(1)0}&:=&r_{(1)}(0) =r-1 \,, \label{BCreq}\\
 z_{(1)0}&:=&z_{(1)}(0)=0 \,, \label{BCzeq} \\
 \Theta_{(1)0} &:=& \Theta_{(1)}(0)=0\,. \label{BCThetaeq}
\end{eqnarray}
\end{subequations}
The first BC (\ref{BCreq}) defines the change in the equatorial radius, $r_{(1)0}$, which will be used as the small parameter characterizing the small deformation; $r_{(1)0}<0$ corresponds to compression, whereas $r_{(1)0}>0$ to dilation. For $C_{(0)}\neq 2$ we can express $r_{(1)0}$ in terms of the value of $\kappa_{(1)}$ at the equator and $C_{(1)}$. Evaluation of Eqs. (\ref{sphedTheta}) and (\ref{DETheta1p}) at $\ell=0$ yields
\begin{subequations}
\begin{eqnarray} \label{BC:r10}
 r_{(1)0} &=& \dot{\Theta}_{(1)0} - \kappa_{(1)0} \,, \\
  (2-C_{(0)})\dot{\Theta}_{(1)0} &=& (1-C_{(0)}) \kappa_{(1)0} + C_{(1)}\,.
\end{eqnarray}
\end{subequations}
Combining both equations to eliminate $\dot{\Theta}_{(1)0}$, we obtain $r_{(1)0}$ in terms of $\kappa_{(1)0}$
\begin{equation} \label{r1cond}
 \left(2-C_{(0)}\right) r_{(1)0} = -\kappa_{(1)0} + C_{(1)}\,.
\end{equation}
The second BC (\ref{BCzeq}) ensures that the deformation preserves the equatorial mirror symmetry. The third BC (\ref{BCThetaeq}) ensures the existence of a tangential plane at the equator.
\\
The fourth BC would correspond to the condition of the existence of a tangent plane at the pole, which to first order implies that $\Theta_{(1)p}=0$, but as mentioned above, it would require also that $r_{(1)p}=0$, completing the five required conditions, which in general does not allow for the fulfillment of the area or volume constraints.
\\
Thus, the second possibility is chosen, which provides the fourth condition
\begin{equation} \label{BCrThetapole}
 r_{(1)p}=-\Theta_{(1)p} \,.
\end{equation}
The vanishing of the numerator entails the regularity of $c_\parallel$ (and of $h_B$) at the pole: using the L'Hopital rule in Eq. (\ref{Eqs:frstordcprpcpar}), we get
\begin{equation}
 \lim_{\ell\to \pi/2} c_{\parallel (1)} = \frac{\dot{\Theta}_{(1)}+ \dot{r}_{(1)}}{\sin \ell}\Big{|}_{\ell = \pi/2} = \dot{\Theta}_{(1)p}\,,
\end{equation}
where in the last step we used Eq. (\ref{sphedr}), which implies that $\dot{r}_{(1)}(\pi/2)=0$. Therefore $c_{\parallel (1)p} = c_{\perp (1)p}$, so the vesicle is umbilical at the pole, and the first-order correction to the mean curvature difference is
\begin{equation} \label{Eq:kappa1p}
\kappa_{(1)p}=k_{(1)p}= 2 \dot{\Theta}_{(1)p} \,.
\end{equation}
The fifth condition is obtained from the constraints of fixed total area or volume. The first-order changes in the total area and volume are
\begin{subequations}
 \begin{eqnarray}
 a_{(1)} &=& \int_0^{\pi/2} r_{(1)} \, \rmd \ell \,, \label{Eq:a1}\\
 v_{(1)} &=& \frac{3}{2}\int_0^{\pi/2} \cos^2 \ell \left(2 r_{(1)} - \sin \ell \Theta_{(1)}  \right) \, \rmd\ell \,. \label{Eq:v1}
\end{eqnarray}
\end{subequations}
Using Eq. (\ref{sphedr}) in Eq. (\ref{Eq:v1}) to replace $\Theta_{(1)}$ in favor of $r_{(1)}$, as well as the double angle formulas, the integrand in Eq. (\ref{Eq:v1}) can be recast as
\begin{equation}
\cos^2 \ell \left(2 r_{(1)} - \sin \ell \Theta_{(1)}  \right) = \frac{1}{2}\frac{\rmd}{\rmd\ell} \left( \sin 2\ell \,r_{(1)}\right) +r_{(1)}\,.
\end{equation}
Since $\sin 2\ell$ vanishes at the boundaries, the first-order change in the total scaled volume is proportional to the change in the total scaled area
\begin{equation}
v_{(1)} =\frac{3}{2}\,a_{(1)} \,.
\end{equation}
Thus, to first order the fulfillment of one constraint also implies the fulfillment of the other one, which requires that
\begin{equation} \label{Eq:a1kappa1perpr0}
a_{(1)} =  \int_0^{\pi/2} r_{(1)} \rmd \ell =0 \,.
\end{equation}
This constraint can be expressed as an orthogonality condition for $\kappa_{(1)}$. To this end, we rewrite Eq. (\ref{sphedTheta}) as a second-order differential equation for $r_{(1)}$
\begin{equation}
\ddot{r}_{(1)} + r_{(1)} + \cos \ell \kappa_{(1)} =0 \,.
\end{equation}
Integrating this equation in the interval $[0,\pi/2]$ and using the fact that $\dot{r}_{(1)}$ vanishes at $\ell=0,\pi/2$ on account of the BC (\ref{BCThetaeq}), we get that the area constraint, Eq. (\ref{Eq:a1kappa1perpr0}), can be recast as
\begin{equation} \label{a1cond}
 a_{(1)} = -\int_0^{\pi/2} \cos \ell \kappa_{(1)} \, \rmd \ell = 0\,.
\end{equation}
This is the last condition required to determine one of the constants of integration or the Lagrange multipliers. It also implies that the total curvature of the vesicle does not change to first order, using Eq. (\ref{Eq:kG1}), we get
\begin{equation}
\int k_{G(1)} \rmd a = 2 \pi \int_0^{\pi/2} \cos \ell \kappa_{(1)} \, \rmd \ell =0\,.
\end{equation}
The first-order correction to the total bending energy is given by
\begin{equation}
h_{B(1)} = \frac{\kappa_{(0)}}{2} \int_0^{\pi/2} \left(  r_{(0)} \kappa_{(1)} + \frac{\kappa_{(0)}}{2} r_{(1)} \right) d \ell \,. 
\end{equation}
Using Eq. (\ref{sphedTheta}) to replace $\kappa_{(1)}$ in favor of $\Theta_{(1)}$ and its first derivative, taking into account the BC (\ref{BCThetaeq}) and the fixed area constraint, Eq. (\ref{Eq:a1kappa1perpr0}), we get that the bending energy does not change to first order 
\begin{eqnarray}
h_{B(1)} &=& \frac{\kappa_{(0)}}{2} \int_0^{\pi/2}  \left[\frac{\rmd}{\rmd \ell}\left( \cos \ell \Theta_{(1)} \right) + \left(\frac{\kappa_{(0)}}{2} -1 \right)  r_{(1)} \right] \rmd \ell  \,, \nonumber \\
&=& \frac{\kappa_{(0)}}{2} \left[ \cos \ell \Theta_{(1)} \Big|_0^{\pi/2} + \left(\frac{\kappa_{(0)}}{2} -1 \right)  a_{(1)} \right] =0 \,.
\end{eqnarray}
Since $\ddot{\Theta}_{(0)}(0)=0$, the zeroth order of the total force $f_0$, given in Eq. (\ref{Eq:f0}), vanishes on the initial spherical vesicle. Taking this into account and that $r_{(0)}(0)=1$, we get that the total force is given by first-order correction
\begin{equation}
\frac{f_{0(1)}}{4 \pi} = - \left(r_{(0)} \ddot{\Theta}_{(1)} + r_{(1)} \ddot{\Theta}_{(0)}\right)\Big|_{\ell=0} = - \ddot{\Theta}_{(1)0}\,.
\end{equation}

\section{Weak deformation of the sphere} \label{SectWeakConstSphere}

Here, we present the solutions of the system of linearized differential equations. First, we consider the cases with constant values $C_{(0)}=0,2$ and then the general case with $C_{(0)}\neq 0,2$.

\subsection{Solution for $C_{(0)}=0$}

For $C_{(0)}=0$ the solution of Eq. (\ref{Eq:ddkappa1}) is
\begin{equation}
 \kappa_{(1)} =  D_{(1)} \ln(\sec \ell + \tan \ell) -2 C_{(1)} \ln \cos \ell + \kappa_{(1)0} \,,
\end{equation}
where $D_{(1)}$ and $\kappa_{(1)0}$ are constants of integration, the latter corresponding to the value of $\kappa_{(1)}$ at the equator.
\\
Since the first-order correction to the total energy is proportional to $\kappa_{(1)}$, in order to avoid a divergence of $\kappa_{(1)}$ at the poles with $\ell = \pi/2$ due to the terms proportional to $\ln \cos \ell$, we set $D_{(1)} =-2 C_{(1)}$, so
\begin{equation}
 \kappa_{(1)} = - 2 C_{(1)} \ln(1 + \sin \ell) + \kappa_{(1)0} \,.
\end{equation}
From Eqs. (\ref{r1cond}) and (\ref{a1cond}) we get the following Eqs.
\begin{subequations}
 \begin{eqnarray}
C_{(1)}-\kappa_{(1)0} &=& 2 \, r_{(1)0} \,,\\
2 C_{(1)} (2 \ln 2-1) - \kappa_{(1)0} &=& 0 \,.
 \end{eqnarray}
\end{subequations}
Solving these equations we determine the constants $C_{(1)}$ and $\kappa_{(1)0}$ in terms of the small parameter $r_{(1)0}$
\begin{subequations}
\begin{eqnarray}
  C_{(1)}&=& A r_{(1)0} \,,\\
  \kappa_{(1)0} &=& \left(A-2\right) r_{(1)0} \,,
  \end{eqnarray}
 \end{subequations}
where we have defined the constant
 \begin{equation}
A= \frac{2}{3-4 \ln 2}  \approx 8.795 \,.
 \end{equation}
Thus
\begin{equation}
 \kappa_{(1)} = \left[- 2 A \ln(1 + \sin \ell) +A-2 \right] r_{(1)0} \,.
\end{equation}
Substituting this result in Eq. (\ref{DETheta1p}) with $C_{(0)}=0$, we get
\begin{eqnarray} \label{sphTheta1p}
 \frac{\dot{\Theta}_{(1)}}{r_{(1)0}} &=& A \left[ \frac{1}{1+\sin \ell} -\ln \left(1+ \sin \ell  \right)  \right] -1 \,.
\end{eqnarray}
Since the integral of $\ln (1+\sin \ell)$ cannot be expressed in closed form, in order to integrate this equation we have resort to its trigonometric expansion, given by
\begin{eqnarray}
 \ln(1+\sin \ell) &=& 2 \displaystyle\sum_{k=1}^\infty (-1)^{k-1} \left[\frac{ \sin (2 k - 1) \ell}{2 k-1} + \frac{ \cos 2 k \ell}{2 k}\right] \nonumber \\
 &&-\ln 2 \,.
\end{eqnarray}
Integrating Eq. (\ref{sphTheta1p}) and imposing the BC (\ref{BCThetaeq}) we obtain
\begin{eqnarray} \label{Eq:Th1sol}
 \frac{\Theta_{(1)}}{r_{(1)0}} &=& 2 A \displaystyle\sum_{k=1}^\infty (-1)^{k-1} \left[ \frac{ \cos (2 k-1)  \ell}{(2 k-1)^2} - \frac{ \sin 2 k \ell}{4 k^2} \right] \nonumber \\
 &-& A \left[\frac{ \cos \ell}{1+ \sin \ell} +2G-1 \right] +\left( A \ln 2-1\right) \ell  \,, \qquad
\end{eqnarray}
where $G$ is Catalan's constant, defined by
\begin{equation}
G = \displaystyle\sum_{k=0}^\infty \frac{(-1)^k}{(2 k + 1)^2} \approx 0.916 \,.
\end{equation}
\vskip0pc \noindent
Substituting the expression for $\Theta_{(1)}$ given in Eq. (\ref{Eq:Th1sol}) in Eqs. (\ref{sphedr}) and (\ref{sphedz}), integrating and imposing BCs (\ref{BCrThetapole}) and (\ref{BCzeq}), we obtain the first-order corrections to the radial and height coordinates in terms of $r_{(1)0}$
\begin{subequations}
\begin{eqnarray}
\frac{r_{(1)}}{r_{(1)0}} &=& A \displaystyle\sum_{k=1}^\infty (-1)^k \left[ \frac{4  \sin 2 k \ell}{(4k^2-1)^2} + \frac{ \cos (2 k+1) \ell}{4 k^2(k+1)^2} \right] \nonumber \\
&+&\left[\left(1-A \ln 2\right) \ell +A (2G-1) \right] \sin \ell \nonumber \\
&+&\left[1-A \left(\ln 2 -\frac{3}{4}\right)\right]  \cos \ell \,,  \\
\frac{z_{(1)}}{r_{(1)0}} &=& A \displaystyle\sum_{k=1}^\infty (-1)^{k+1} \frac{(4k^2+1)}{ k (4k^2-1)^2}  \left(\cos 2 k \ell -1\right) \nonumber \\
&+&A \displaystyle\sum_{k=1}^\infty \frac{(-1)^k}{4 (2 k+1)} \left( \frac{1}{k^2} + \frac{1}{(k+1)^2} \right) \sin (2 k+1) \ell \nonumber \\
&+&\left(1+A\left(\frac{5}{4}-\ln 2 \right) \right) \sin \ell -A \ln (1+\sin \ell)  \nonumber \\
&+& \left( A \ln 2 -1 \right) \ell  \cos \ell \nonumber \\
&+& A \left( 2 G-1\right)  \left(1-\cos \ell \right) \,. 
\end{eqnarray}
\end{subequations}
Differentiating Eq. (\ref{sphTheta1p}) we get that the second derivative of $\Theta_{(1)}$ is given by
\begin{equation} \label{sphTheta1pp}
\frac{\ddot{\Theta}_{(1)}}{r_{(1)0}} = - \frac{A \cos \ell}{1+\sin \ell} \left( \frac{1}{1+\sin \ell} + 1 \right)  \,.
\end{equation}
Evaluating Eqs. (\ref{sphTheta1p}) and (\ref{sphTheta1pp}) at the equator, we get that  $\dot{\Theta}_{(1)0}$ and $f_{0(1)}$ are given by \begin{subequations}
\begin{eqnarray}
 \dot{\Theta}_{(1)0} &=& (A - 1) r_{(1)0}  \approx 7.795 \, r_{(1)0}\,,  \\
 \frac{f_{0(1)}}{4 \pi} &=& -\ddot{\Theta}_{(1)0} =  2 A r_{(1)0} \approx 17.589 \, r_{(1)0}  \,.
\end{eqnarray}
\end{subequations}
At the pole the values of the first-order corrections of $r$ and $\Theta$ are
\begin{eqnarray} \label{r1pC0eq0}
 r_{(1)p}= -\Theta_{(1)p} &=& \left[\left(1-A \ln 2\right) \frac{\pi}{2} +A (2G-1)\right] r_{(1)0} \nonumber \\
 &\approx& -0.688 r_{(1)0} \,.
\end{eqnarray}

\subsection{Solution for $C_{(0)} = 2$}

For $C_{(0)}=2$ Eq. (\ref{sphedkappa}) becomes an ODE for $\kappa_{(1)}$, whose solution is
\begin{equation} \label{k1C0eq2}
\kappa_{(1)}=A_{(1)} \sin \ell +C_{(1)}\,.
\end{equation}
where $A_{(1)}$ is a constant of integration. In this case, the l.h.s. of Eqs. (\ref{DETheta1p}) and (\ref{DETheta1pddkappa}) vanish, so they cannot be used to determine $\Theta$. Instead, multiplying Eq. (\ref{sphedTheta}) by $\cos \ell$, differentiating it and using Eq. (\ref{sphedr}) to replace $\dot{r}_{(1)}$ in favor of $\Theta_{(1)}$, we get a second-order differential Eq. for $\Theta_{(1)}$
\begin{equation} \label{eq:ddotTheta}
 \cos \ell \ddot{\Theta}_{(1)} - 2 \sin \ell \dot{\Theta}_{(1)} -\cos \ell \dot{\kappa}_{(1)}+ \sin \ell \kappa_{(1)} =0 \,.
\end{equation}
For $\kappa_{(1)}$ given in Eq. (\ref{k1C0eq2}) the general solution of Eq. (\ref{eq:ddotTheta}) is given by
\begin{eqnarray}
\Theta_{(1)} &=& \sec \ell \left( -\frac{1}{3} A_{(1)} \cos 2 \ell + D_{(1)} \sin \ell \right) \nonumber \\&&
+  \frac{C_{(1)}}{2} \ell + E_{(1)}\,,
\end{eqnarray}
where $D_{(1)}$ and $E_{(1)}$ are constants of integration. Imposing the BC at the equator for $\Theta_{(1)}$ given by Eq. (\ref{BCThetaeq}), as well as the regularity of $\Theta_{(1)}$ at the pole with $\ell=\pi/2$, we determine these two constants of integration in terms of $A_{(1)}$
\begin{equation}
D_{(1)} = -\frac{A_{(1)}}{3} \,, \quad E_{(1)} = \frac{A_{(1)}}{3} \,.
\end{equation}
Thus, we get
\begin{equation}
\Theta_{(1)} =  \frac{A_{(1)}}{3} \left[\frac{\cos \ell}{1 + \sin \ell} -2 \cos \ell  +1 \right] 
+  \frac{C_{(1)}}{2} \ell \,,
\end{equation}
Using these results, integrating Eq. (\ref{sphedr}) and imposing the BC (\ref{BCrThetapole}) we get the general solution
\begin{eqnarray}
r_{(1)} &=& \frac{A_{(1)}}{3} \left[\frac{\sin 2 \ell}{2}- \sin \ell - \cos \ell \right]\nonumber \\
&-&\frac{C_{(1)}}{2} \left(\ell \sin \ell + \cos \ell \right) \,.
\end{eqnarray}
Imposing the BC (\ref{BCreq}) and the area constraint (\ref{Eq:a1kappa1perpr0}) we determine both constants in terms of $r_{(1)0}$
\begin{equation}
 A_{(1)} = -12 \, r_{(1)0} \,, \quad C_{(1)} = 6 \, r_{(1)0} \,.
\end{equation}
Using these results and integrating also (\ref{sphedz}) with the BC (\ref{BCzeq}), we get the radial and height coordinates, as well as $\Theta_{(1)}$ in terms of $r_{(1)0}$
\begin{subequations}
\begin{eqnarray}
 \frac{r_{(1)}}{r_{(1)0}}&=&\cos \ell - \sin \ell \left( 4 \cos \ell + 3 \ell -4\right) , \\
\frac{z_{(1)}}{r_{(1)0}}&=&\sin \ell + \cos \ell \left( 4 \cos \ell +3 \ell- 4 \right)\nonumber \\
&-&4 \ln (1+ \sin \ell)\,.
\end{eqnarray}
\end{subequations}
The first order correction of $\Theta$ and its derivatives are given by
\begin{subequations}
\begin{eqnarray}
 \frac{\Theta_{(1)}}{r_{(1)0}}&=& \frac{2 \sin 2\ell}{1+\sin \ell}+ 4 \cos \ell  +3 \ell-4 \,, \\
\frac{\dot{\Theta}_{(1)}}{r_{(1)0}}&=& \frac{4 ( \cos 2 \ell+2)}{1+\sin \ell}-5 \,, \\
\frac{\ddot{\Theta}_{(1)}}{r_{(1)0}}&=& \frac{2 (\cos 3 \ell-4 \sin 2 \ell -7 \cos \ell )}{(1+\sin \ell)^2} \,.
\end{eqnarray}
 \end{subequations}
Evaluating the derivatives at the equator, we get that the initial value of the derivative of $\Theta_{(1)}$ and the first-order correction to the force are
\begin{subequations}
\begin{eqnarray}
\dot{\Theta}_{(1)0} &=& 7 \, r_{(1)0}\,, \\
\frac{f_{0(1)}}{4 \pi} &=& -\ddot{\Theta}_{(1)0}= 12 \, r_{(1)0} \,.
\end{eqnarray}
\end{subequations}
 At the pole the corrections of $r$ and $\Theta$ are
\begin{equation}
 r_{(1)p} =-\Theta_{(1)p} = \left(4 - \frac{3 \pi}{2}\right) r_{(1)0} \approx -0.712 \, r_{(1)0} \,.
\end{equation}
 
\subsection{Solution for $C_{(0)} \neq 0,2$}

\begin{figure}[t]
\begin{center}
\begin{tabular}{cc}
  $\vcenter{\hbox{\includegraphics[width = 0.45 \linewidth]{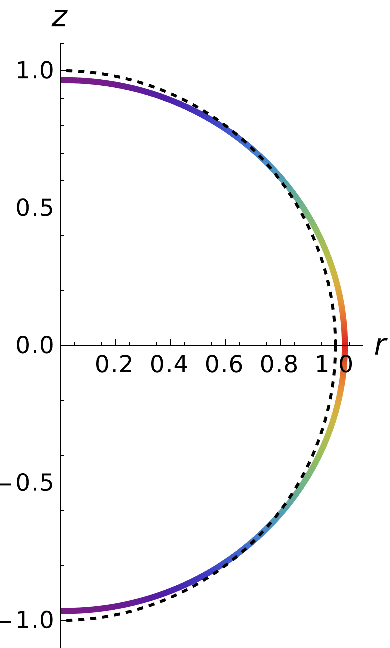}}}$ &
  $\vcenter{\hbox{\includegraphics[width = 0.45\linewidth]{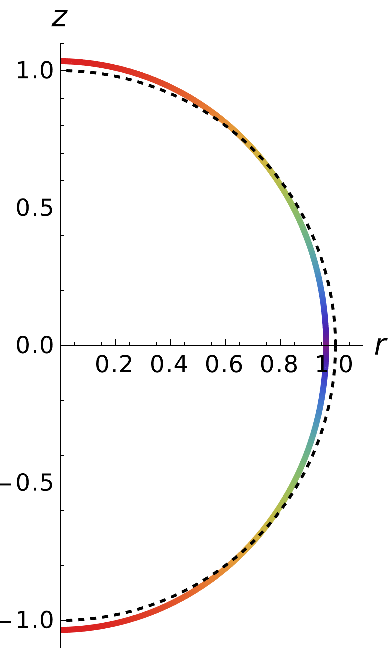}}}$ \\
  {\small (a) $r_{(1)0}>0$} & {\small (b) $r_{(1)0}<0$}
  \end{tabular}
\includegraphics[scale=0.75]{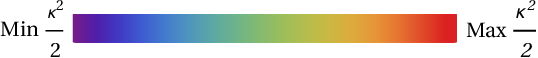}
  \end{center}
\caption{ (a) Dilation and (b) constriction of a spherical vesicle with parameter $C_{(0)}=1$. The initial spherical configuration is shown with a black dashed line. The magnitude of the deformations has been augmented for illustration purposes. The bending energy density is color coded.}
\label{Fig:2}
\end{figure}

\begin{figure}
  \includegraphics[width = \linewidth]{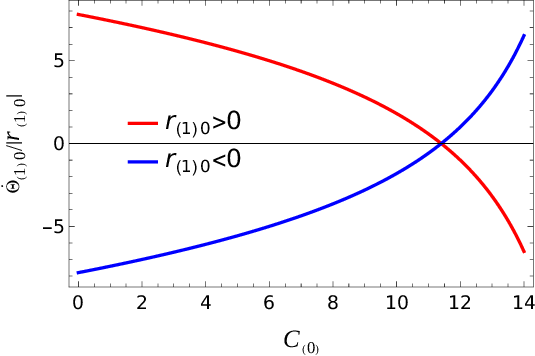}
\caption{Values of $\dot{\Theta}_{1}$ at the equator as a function of $C_{(0)}$. The red (blue) line corresponds to dilation (constriction) of the membrane, with $r_{(1)0}>0$ ($r_{(1)0}<0$).}
\label{Fig:3}
\end{figure}

\begin{figure}
  \includegraphics[width= \linewidth]{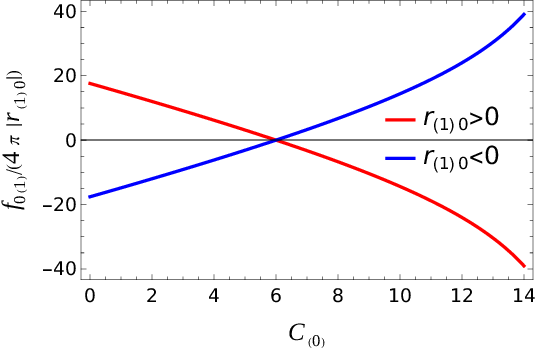}
\caption{Total force as a function of $C_{(0)}$. The red (blue) line corresponds to dilation (constriction) of the membrane, with $r_{(1)0}>0$ ($r_{(1)0}<0$).}
\label{Fig:4}
\end{figure}

\begin{figure}
  \includegraphics[width= \linewidth]{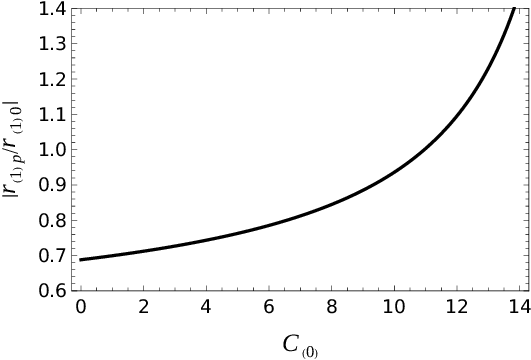}
\caption{Magnitude of the first-order correction to the radial coordinate at the pole as a function of the parameter $C_{(0)}$.}
\label{Fig:5}
\end{figure}

For $C_{(0)} \neq 0$, Eq. (\ref{Eq:ddkappa1}) can be recast as
\begin{equation}
\frac{d}{d \ell} \left(\cos \ell \dot{\kappa}_{(1)} \right) +  C_{(0)} \cos \ell \left( \kappa_{(1)} -2 \frac{C_{(1)}}{C_{(0)}}  \right) =0 \,.
\end{equation}
By making the change of variable, coordinate and parameter
\begin{subequations}
\begin{eqnarray}
u_{(1)}&=&\kappa_{(1)}-2\frac{C_{(1)}}{C_{(0)}}\,, \\
\theta&=&\frac{\pi}{2}-\ell\,, \\
\nu(\nu+1)&=&C_{(0)}\,, \quad \nu \neq -1,0 \,,
 \end{eqnarray}
\end{subequations}
Eq. (\ref{Eq:ddkappa1}) can be recast as the Legendre equation
\begin{equation}
 \frac{1}{\sin \theta} \frac{d}{d\theta} \left(\sin \theta \frac{d}{d \theta}u_{(1)}(\theta)\right) + \nu (\nu +1)  u_{(1)}(\theta)=0\,.
 \end{equation}
The solution is given by
\begin{eqnarray}
 u_{(1)}&=&A_{(1)} P_{\nu}(\cos \theta)+B_{(1)} Q_{\nu}(\cos \theta ) \,, \\
 \nu&=&\frac{1}{2} \left(\pm \sqrt{1+ 4 C_{(0)}}-1\right)\,, \nonumber
\end{eqnarray}
where $P_\nu(x)$ and $Q_\nu(x)$ are the Legendre functions of the first and second kind \cite{AbramowitzBook, ArfkenBook, Gradshteyn2007}, and $A_{(1)}$ and $B_{(1)}$ are constants of integration. Although we could consider the case where the parameter $\nu$ becomes complex, for the sake of concreteness we consider that $C_{(0)}>-1/4$, such that $\nu \in \mathbbm{R}$.\footnote{For $C_{(0)} < -1/4$ the parameter $\nu$ becomes complex,  but the Legendre functions $P_\nu(\cos \theta)$ with complex parameter $\nu=-1/2+ i a$, known as conical functions, are real for $\theta \in \mathbbm{R}$ \cite{Gradshteyn2007}.} Since $Q_\nu(\sin \ell) \rightarrow \infty$ as $\ell \rightarrow \pi/2$, to avoid a singularity at the poles we set $B_{(1)}=0$. Thus, reverting to the original parameter, we have
\begin{equation} \label{kappa1simp}
 \kappa_{(1)} = A_{(1)} P_{\nu}(\sin \ell)+\frac{2 C_{(1)}}{C_{(0)}} \,.
 \end{equation}
In this case, the BC (\ref{r1cond}) and the area constraint (\ref{a1cond}) read
\begin{subequations}
\begin{eqnarray}
A_{(1)}\zeta_\nu + \frac{C_{(1)}}{C_{(0)}}  &=& - r_{(1)0} \,, \\
A_{(1)}\xi_\nu -\frac{C_{(1)}}{C_{(0)}} &=& 0 \,,
\end{eqnarray}
\end{subequations}
where we have defined
\begin{subequations}
\begin{eqnarray}
\zeta_\nu&=&\frac{P_\nu(0)}{2-\nu (\nu+1)}\,, \\
\xi_\nu &=& \frac{P_{\nu+1}(0) -P_{\nu-1}(0)}{2 (2 \nu + 1)} \,,  \\
 P_\nu(0) &=& \frac{\sqrt{\pi}}{\Gamma\left(1+\frac{\nu}{2}\right) \Gamma\left(\frac{1-\nu}{2}\right)} \,. \nonumber
\end{eqnarray}
\end{subequations}
At first glance $\zeta_\nu$ is indeterminate as $\nu \rightarrow 1$, but taking the limit appropriately, we get that $\zeta_\nu \rightarrow 1/3$.
\\
Solving these equations we determine $A_{(1)}$ and $C_{(1)}$ in terms of $r_{(1)0}$
\begin{subequations}
\begin{eqnarray}
A_{(1)} &=& -\frac{r_{(1)0}}{\zeta_\nu+\xi_\nu}  \,, \\
\frac{C_{(1)}}{C_{(0)}} &=&  -  \frac{\xi_\nu \, r_{(1)0}}{\zeta_\nu + \xi_\nu} \,.
\end{eqnarray}
\end{subequations}
Thus, the first-order approximation is valid only for values of $\nu$ (or $C_{(0)}$) for which $\zeta_\nu+\xi_\nu\neq0$. The first roots of the equation $\zeta_\nu+\xi_\nu=0$ occur at $\nu=-1,0,3.686$.
\\
Having determined $A_{(1)}$ and $C_{(1)}$, we can  express $\kappa_{(1)}$ and its derivative in terms of $r_{(1)0}$ as
\begin{subequations}
\begin{eqnarray}
\frac{\kappa_{(1)}}{r_{(1)0} } &=& -\left[\frac{ P_{\nu }(\sin \ell)+ 2 \xi_\nu}{\zeta_\nu + \xi_\nu} \right] \,, \label{Eq:kappa1r10} \\
 \frac{\dot{\kappa}_{(1)}}{r_{(1)0} } &=& \frac{\nu +1}{\zeta_\nu+\xi_\nu} \sec \ell \left[P_{\nu +1}(\sin \ell)-\sin \ell \, P_{\nu }(\sin \ell) \right], \qquad  \label{Eq:dkappa1r10}
\end{eqnarray}
 \end{subequations}
Integrating Eq. (\ref{DETheta1pddkappa}) we can determine $\Theta_{(1)}$. However, this involves the integral of $P_\nu(\sin \ell)$, which cannot be expressed in closed form, so we use its trigonometric expansion, (a derivation can be found in Appendix \ref{AppB})
\begin{align}
&P_\nu (\sin \ell) = \displaystyle\sum_{n=1}^{\infty} \displaystyle\sum_{k=1}^n \left[\alpha^n_k \cos 2 k \ell + \beta^n_k \sin (2 k - 1) \ell \right]\nonumber \\
&\qquad \quad + P_{\nu}(0)^2\,, \label{Eq:trigexpPnsinl} \\
&\alpha^n_{k,\nu} = \frac{2 \sqrt{\pi}(-1)^k}{\Gamma\left(1+\frac{\nu}{2}-n\right) \Gamma\left(\frac{1-\nu}{2}-n\right) (n-k)! (n+k)!} \,, \nonumber  \\
&\beta^n_{k,\nu} = \frac{2 \sqrt{\pi}(-1)^k}{\Gamma\left(\frac{3+\nu}{2}-n\right) \Gamma\left(1-\frac{\nu}{2}-n\right) (n-k)! (n+k-1)!} \,. \nonumber
\end{align}
Note that the constant term in Eq. (\ref{Eq:trigexpPnsinl}) is $P_{\nu}(0)^2$, not $P_{\nu}(0)$, which is due to the fact that
\begin{equation}
 \alpha_\nu : = \displaystyle\sum_{n=1}^{\infty} \displaystyle\sum_{k=1}^n \alpha^n_{k,\nu} = P_\nu(0)-P_{\nu}(0)^2\,.
\end{equation}
Using these results and integrating Eq. (\ref{DETheta1pddkappa}) with the BCs for $\Theta_{(1)}$ given in Eq. (\ref{BCThetaeq}), we get
\begin{eqnarray} \label{Theta1C0neq0initial}
\frac{\Theta_{(1)}}{r_{(1)0}} &=& \upsilon_\nu \Big[ \displaystyle \sum_{n=1}^\infty\sum_{k=1}^n \left(2 k -\frac{1}{2k}\right) \alpha^n_{k,\nu}  \sin 2k \ell \nonumber \\
&+& \displaystyle \sum_{n=2}^\infty\sum_{k=2}^n \left( \frac{1}{2k-1} -2k+1 \right) \beta^n_{k,\nu}  \cos (2k-1) \ell \nonumber \\
&-& \chi_\nu \, \ell - \eta_\nu \Big] \,,
\end{eqnarray}
where we have defined
\begin{subequations}
\begin{eqnarray}
\upsilon_\nu &=& \frac{1}{(2-\nu (\nu+1)) (\zeta_\nu + \xi_\nu)} \,, \\
\chi_\nu &=& P_\nu(0)^2 + \left(2-\nu (\nu+1) \right) \xi_\nu \,, \\
\eta_\nu &= & \displaystyle \sum_{n=2}^\infty\sum_{k=2}^n \left(\frac{1}{2k-1} -2k+1 \right) \beta^n_{k,\nu} \,.
\end{eqnarray}
\end{subequations}
In order to avoid that $\upsilon_\nu$ diverges as $\nu \rightarrow 1$, it is also required that $\nu \neq 1$ or $C_{(0)} \neq 2$.  Using Eqs. (\ref{IntsI}) of Appendix \ref{AppA} we can integrate  Eqs. (\ref{sphedr}) and (\ref{sphedz}) with the BCs (\ref{BCrThetapole}) and (\ref{BCzeq}) we obtain the radial and height coordinates
\begin{subequations}
\begin{eqnarray}
&&\frac{r_{(1)}}{r_{(1)0}} = \upsilon_\nu \Big[ \displaystyle \sum_{n=1}^\infty\sum_{k=1}^n \frac{\alpha^n_{k,\nu}}{4k} \times  \nonumber \\
&&\left[ (2k+1) \cos (2k-1) \ell +  (2k-1) \cos (2k+1) \ell \right] \nonumber \\
&&+ \displaystyle \sum_{n=2}^\infty\sum_{k=2}^n \frac{\beta^n_{k,\nu}}{2k-1} \left[ k \sin 2(k-1) \ell + (k-1) \sin 2k \ell \right] \nonumber \\
&&+ \chi_\nu \left( \ell \sin \ell + \cos \ell \right) + \eta_\nu \sin \ell \Big]\,, \\
&&  \frac{z_{(1)}}{r_{(1)0}} = \upsilon_\nu \Big[\displaystyle \sum_{n=1}^\infty \sum_{k=1}^n \frac{\alpha^n_{k, \nu}}{4k} \times \nonumber \\
 &&\left[(2k-1) \sin (2k+1) \ell - (2k+1) \sin (2k-1) \ell \right]  \nonumber \\
&&+ \displaystyle \sum_{n=2}^\infty\sum_{k=2}^n \frac{\beta^n_{k,\nu} }{2k-1} \left[ k (\cos 2(k-1) \ell  -\cos 2k \ell) -2 \sin^2 k \ell \right] \nonumber \\
&&-\chi_\nu \left( \ell \cos \ell - \sin \ell  \right) + 2 \, \eta_\nu \, \sin^2 \frac{\ell}{2}  \Big] \,.
\end{eqnarray}
\end{subequations}
The initial deformation of a spherical membrane given by these first order corrections of the radial and height coordinates is illustrated in Fig. \ref{Fig:2} for dilation ($r_{(1)0}>0$) and compression ($r_{(1)0}<0$). The scaled bending energy density $\kappa^2/2$ is color coded in these figures, where we see that for dilation (constriction) it is maximum (minimum) at the equator, whereas it is minimum (maximum) at the poles, which agrees with the numerical results of Ref. \cite{Vazquez2025}.
\vskip0pc \noindent
Evaluating Eq. (\ref{DETheta1p}) at the equator we get that the initial value of the first derivative of the first-order correction to the angle $\Theta$ is
\begin{equation} \label{dTheta1C0neq0}
\frac{\dot{\Theta}_{(1)0}}{r_{(1)0}} = \frac{\nu (\nu+1) \zeta_\nu}{\zeta_\nu + \xi_\nu} -1 \,,
\end{equation}
which is plotted in Fig. \ref{Fig:3} as a function of the parameter $C_{(0)}$. For dilation (compression), corresponding to $r_{(1)0}>0$ ($r_{(1)0}<0$), $\dot{\Theta}_{(1)0}$ decreases (increases) with $C_{(0)}$, and it vanishes at $C_{(0)}=11.4$. 
\vskip0pc \noindent
Differentiating Eq. (\ref{DETheta1p}) and using Eq. (\ref{DETheta1pddkappa}) to replace $ \ddot{\kappa}_{(1)}$ in favor of $\kappa_{1}$ and $\dot{\kappa}_{(1)}$ we obtain the second derivative of $\Theta_{(1)}$
\begin{eqnarray}
 (2-C_{(0)})\ddot{\Theta}_{(1)} &=& (2 \sec^2 \ell -C_{(0)}) \dot{\kappa}_{(1)} \nonumber \\
 &-& \tan \ell \left(C_{(0)} \kappa_{(1)} -2C_{(1)} \right) \,.
\end{eqnarray}
Evaluating this result at the equator we obtain $\ddot{\Theta}_{(1)0}=\dot{\kappa}_{(1)0}$. Using Eq. (\ref{Eq:dkappa1r10}) we get that the first-order correction to the total force is
\begin{equation}  \label{f10C0neq0}
\frac{f_{(1)0}}{4 \pi} =-\ddot{\Theta}_{(1)0} 
= - \frac{(\nu+1)}{\zeta_\nu + \xi_\nu} \, P_{\nu+1}(0) r_{(1)0} \,,
\end{equation}
which is plotted in Fig. \ref{Fig:4} as a function of the parameter $C_{(0)}$. For $C_{(0)}<6$, the direction of $f_{(1)0}$ correlates with the direction of the deformation of the membrane: an inward (outward) force $f_{0(1)}<0$ ($f_{0(1)}>0$) is required to constrict (dilate) the membrane, for which $r_{(1)0}<0$ ($r_{(1)0}>0$). At $C_{(0)}=6$ the total force vanishes. For $C_{(0)}>6$ the total force changes sign, so its direction anticorrelates with the direction of the deformation of the membrane, an outward (inward) force $f_{0(1)}>0$ ($f_{0(1)}<0$) is required to constrict (dilate) the membrane.
\vskip0pc \noindent
Despite the fact that $\dot{\Theta}_{(0)}$ and $f_{(1)0}$ are proportional to $(\zeta_\nu+\xi_\nu)^{-1}$, which diverges for $\nu=-1,0$ or $C_{(0)}=0$, as shown in Figs. \ref{Fig:2} and \ref{Fig:3}, their limits agree with the values that we obtained before for that case and for $C_{(0)}=2$.
\\
For comparison purposes, the analytical data of the initial values $\dot{\Theta}_{0(1)}$ and $\ddot{\Theta}_{0(1)}$ (rounded to three decimal places), given by Eqs. (\ref{dTheta1C0neq0}) and (\ref{f10C0neq0}), are displayed in Table \ref{TableI}  for several values of $C_{(0)}$ (rather than of $\nu$), whereas the numerical data  corresponding to the cases of fixed area and fixed volume, considered in Ref. \cite{Vazquez2025}, are shown in Tables \ref{TableII} and \ref{TableIII}, respectively. For the case of constant area with $a=1$, the volume varies, so $\mrmp_{(0)}=0$, and from Eq. (\ref{def:C0}) we have that the parameter is equal to the scaled spontaneous curvature, $C_{(0)}=c_s$. By contrast, for the case of constant volume with $v=1$, the area varies, so $\mu_{(0)}=0$, and the parameter is quadratic in the scaled spontaneous curvature $C_{(0)}=c_s \left(2-c_s/2\right)$, so it is nonnegative only in the interval $c_s \in [0,4]$, for which $C_{(0)} \in [0,2]$.\footnote{We consider only values in this interval, because as mentioned above, for negative values of $C_{(0)}$ the parameter $\nu$ becomes imaginary and one would have to consider the conical functions, which lies beyond the scope of this paper.} We observe that the analytical data show a good agreement with the numerical data.
\begin{table}
 \begin{tabular}{||c|c|c||}
 \hline \hline
$C_{(0)}=c_s$ & $\dot{\Theta}_{(1)0}/r_{(1)0}$ & $\ddot{\Theta}_{(1)0}/r_{(1)0}$ \\
 \hline \hline
 $0$ & 7.795 & -17.589 \\
 \hline
$1$ & 7.410 & -14.820 \\
\hline
$3/2$ & 7.208 & -13.417 \\
\hline
$2$ & 7  & -12 \\
\hline
$3$ & 6.561  & -9.122 \\
\hline \hline
 \end{tabular}
 \caption{Analytical data of the initial values $\dot{\Theta}_{0(1)}$ and $\ddot{\Theta}_{0(1)}$ for different values of $C_{(0)}$ and $r_{(1)0}>0$.} \label{TableI}
\end{table}
\begin{table}
 \begin{tabular}{||c|c|c||}
 \hline \hline
$C_{(0)}=c_s$ & $\dot{\Theta}_{(1)0}/r_{(1)0}$ & $\ddot{\Theta}_{(1)0}/r_{(1)0}$ \\
 \hline \hline
 $0$ & 7.794 & -17.588 \\
 \hline
$1$ & 7.350 & -14.680 \\
\hline
$2$ & 6.980  & -11.973 \\
\hline
$3$ & 6.561  & -9.123 \\
\hline \hline
 \end{tabular}
 \caption{Numerical data of the initial values $\dot{\Theta}_{0(1)}$ and $\ddot{\Theta}_{0(1)}$ for different values of $C_{(0)}$ and $r_{(1)0}>0$, corresponding to vesicles with constant area, $a=1$.} \label{TableII}
\end{table}
\begin{table}
 \begin{tabular}{||c|c|c|c||}
 \hline \hline
$c_s$ & $C_{(0)}=c_s\left(2-\frac{c_s}{2}\right)$ & $\dot{\Theta}_{(1)0}/r_{(1)0}$ & $\ddot{\Theta}_{(1)0}/r_{(1)0}$\\
 \hline \hline
$0$ &  0& 7.781 & -17.579 \\
 \hline
$1$ & 3/2 & 7.208 & -13.416 \\
\hline
$2$ & 2 & 7.000 & -12.117  \\
\hline
$3$ & 3/2 & 7.208 & -13.417  \\
\hline \hline
 \end{tabular}
 \caption{Numerical data of the initial values $\dot{\Theta}_{0(1)}$ and $\ddot{\Theta}_{0(1)}$ for different values of $C_{(0)}$ and $r_{(1)0}>0$, corresponding to vesicles with constant volume, $v=1$.} \label{TableIII}
\end{table}
\vskip0pc \noindent
The values of $r_{(1)}$ and $\Theta_{(1)}$ at the pole are given by
\begin{equation}
r_{(1)p} =-\Theta_{(1)p} = \upsilon_\nu \left( \frac{\pi}{2} \chi_\nu +\eta_\nu \right) r_{(1)0}\,,
 \end{equation}
plotted in Fig. \ref{Fig:5}. We see that for $C_{(0)}<11$, $|r_{(1)p}/r_{(1)0}|<1$, so the extent by which the vesicle does not close, nor it possesses a common tangent plane at the pole, is smaller than $r_{(1)0}$, which is admissible for the numerical analysis of the membrane configurations.

\section{Discussion and conclusions} \label{SecDisConc}

We studied the onset of the deformation of a spherical vesicle produced by a rigid ring whose radius is slightly smaller or larger than the vesicle radius. Employing a perturbative analysis about the sphere, we obtained analytic solutions of the linearized first integral of the EL equation, which describe the initial stage of the deformation. The small parameter of the perturbative expansion is given by the change in the equatorial radius $r_{(1)0}$, whose sign determines the type of deformation of the membrane: if positive (negative) it corresponds to an equatorial dilation (constriction) so the membrane adopts an oblate (a prolate) shape. The analytic solutions are characterized by the parameter $C_{(0)}$, which involves the scaled spontaneous curvature of the membrane $c_s$, the scaled intrinsic surface tension $\mu_{(0)}$ and the scaled pressure difference $\mrmp_{(0)}$, so most of the configurations are degenerate. We found that the solutions of the general case with $C_{(0)}\neq 0,2$ agree with the solutions of the cases with constant values $C_{(0)}=0,2$.
\\
We considered the set of analytic solutions that satisfy the global constraints of constant area and volume. We found that in the linear regime both constraints are related for vesicles with axial and equatorial mirror symmetries. One of the BCs that we used entails the regularity of the membrane curvature at the poles. Although this choice does not allow for the existence of a common tangent plane at the poles, we quantified the mismatch to be of the order of the small parameter of the perturbative expansion, which is admissible for the numerical analysis. One could consider the alternative case where the condition of the tangent plane at the pole is imposed, but in general such choice does not fulfill the constant area and volume constraints.
\\
Even though the first-order corrections to the bending energy density and to the Gaussian curvature of the spherical vesicle are non-vanishing, we demonstrated that its total bending energy and its total curvature do not change in the linear regime. We also calculated analytically the critical value of the total force that is required to initiate the deformation of a spherical vesicle, akin to the critical value that  of the buckling of rod in the Euler instability \cite{LandauBook, AudolyBook}. We found that the total force changes sign at $C_{(0)}=6$.
\\
The perturbative analysis about the sphere that we developed in this work provides insight in the solutions of the first integral of the EL, which in general is difficult to solve due to its singular behavior at the equator, as well as to the requirement to fulfill the constraints of fixed area and volume, along with the constraint of fixed equatorial radius, imposed by the rigid ring. We found that the singularities and divergences that are encountered when solving numerically the EL equation in the nonlinear regime are present even in the linear regime. The analytical results for the initial values of the derivatives of the angle $\Theta$ show a good agreement with the numerical results of Ref. \cite{Vazquez2025}, so they provide a starting point of the numerical analysis of the configurations in the nonlinear regime, which would be useful in the elaboration of the corresponding phase diagrams of vesicles subjected to external forces. Furthermore, the analytic results for the first-order change of the vesicle configuration, as well as the determination of the force required to initiate such deformation, might be relevant for the description of the initial stage of cellular division processes, such as the cytokinesis.
\\
There are several generalizations and refinements that could be explored in order to extend our results. Here, we considered the membrane deformation by a rigid ring, approximation that describes the physical situation that the elastic energy of the ring is much greater than that of the membrane. One could consider the general case of the coupled deformation of a vesicle by a flexible filament \cite{Vetter2014}, which might give rise to non-axisymmetric configurations even for small deformations. Also, for the sake of simplicity, we assumed that the thickness of the filament is negligible, approximation that is valid if the radius of its cross section is much smaller than the radius of the spherical vesicle. However, although it would be more involved, one could take into account its finiteness and model it as a torus \cite{Biron2005, Bahrami2019}. In such case the region of the membrane in contact with the ring would adopt the shape of the torus, and the determination of the point of detachment would provide the appropriate boundary conditions representing the mechanical equilibrium, which might involve localized forces \cite{Cerda2005}.

\section*{Acknowledgements}

A. S.-C acknowledges a doctoral fellowship granted by SECIHTI. P. V.-M. acknowledges support by SECIHTI under programs Investigadores por México (grant 439-2018) and Sistema Nacional de Investigadores (131142).

\begin{appendix}

\section{List of useful integrals} \label{AppA}

The following integrals are used in the determination of the coordinates of the generating curve of the membrane.
\begin{subequations} \label{IntsI}
\begin{align}
& \int \cos \ell \sin 2 k \ell \, d \ell = \nonumber \\
& -\frac{1}{2} \left[ \frac{\cos (2k-1) \ell}{2k-1} +\frac{\cos (2k+1) \ell}{2k+1} \right] \,,  \\
& \int \cos \ell \cos (2 k-1) \ell \, d \ell = \nonumber \\
& \begin{cases}
\frac{1}{2} \left(\ell+ \frac{\sin 2 \ell}{2} \right) ,  & k=1 \\
\frac{1}{2} \left[ \frac{\sin 2(k-1) \ell}{2(k-1)} +\frac{\sin 2 k \ell}{2k} \right] ,  & k>1
 \end{cases}  \\
&\int \sin \ell \sin 2 k \ell \, d \ell =  \nonumber \\
& \frac{1}{2} \left[ \frac{\sin (2k-1) \ell}{2k-1} -\frac{\sin (2k+1) \ell}{2k+1} \right] \,, \\
&\int \sin \ell \cos (2 k-1) \ell \, d \ell  =  \nonumber \\
&\begin{cases}
-\frac{1}{4} \cos 2 \ell ,  & k=1 \\
\frac{1}{2} \left[\frac{\cos 2 (k-1) \ell}{2(k-1)} - \frac{\cos 2k \ell}{2k} \right] ,  & k>1
\end{cases}
 \end{align}
\end{subequations}
\begin{subequations}  \label{IntsII}
 \begin{eqnarray}
 &&\int \sum_{k=1}^{\infty} (-1)^k \frac{\cos \ell \sin 2k\ell}{4 k^2} \, d\ell = \frac{1}{8}\cos \ell \nonumber
 \\
 &&+  \frac{1}{8} \sum_{k=1}^{\infty} (-1)^{k+1}\frac{\cos (2k+1) \ell}{k^2(k+1)^2} \,,  \\
 &&\int \sum_{k=1}^{\infty} (-1)^{k-1} \frac{\cos \ell \cos (2k-1) \ell}{(2k-1)^2} \, d \ell \nonumber \\
 &&= \frac{1}{2} \left[ \ell + \sum_{k=1}^{\infty}(-1)^{k+1} \frac{4 \sin 2 k \ell}{(4k^2-1)^2} \right] \,, \\
 &&\int \sum_{k=1}^{\infty} (-1)^{k-1} \frac{\sin \ell \sin 2k \ell}{4 k^2} \, d \ell = \frac{1}{8} \sin \ell \nonumber \\
 &&+ \frac{1}{8} \sum_{k=1}^{\infty} \frac{(-1)^{k}}{2k+1}\left[\frac{1}{k^2}+\frac{1}{(k+1)^2}\right]\sin (2k+1) \ell  \,, \quad  \\
 &&\int \sum_{k=1}^{\infty} (-1)^k \frac{\sin \ell \cos (2k-1) \ell}{(2k-1)^2} \, d \ell \nonumber \\
 &&=  \sum_{k=1}^{\infty}(-1)^{k+1} \frac{(4k^2+1) \cos 2 k \ell}{2 k (4k^2-1)^2}  \,.
\end{eqnarray}
\end{subequations}

\section{Derivation of the trigonometric expasion of $P_\nu(\sin \ell)$} \label{AppB}

Using the formula of the associated Legendre functions $P^\mu_\nu(x)$ in terms of the derivatives of the Legendre functions $P_\nu(x)$ \cite{Gradshteyn2007}
\begin{equation}
P^n_\nu(x) = (-1)^n (1-x^2)^{n/2} \frac{d^n}{dx^n} P_\nu(x)\,,
\end{equation}
we obtain that the Taylor series of $P_\nu(x)$ at $x=0$, is given by
\begin{equation} \label{TayloSerPnu}
P_\nu (x) = \displaystyle\sum_{n=0}^\infty \frac{(-1)^n}{n!} P_\nu^n(0) x^n \,,
\end{equation}
where 
\begin{equation} \label{Eq:Pnuxeq0}
 P_\nu^n(0) = \frac{2^n \sqrt{\pi}}{\Gamma(\frac{\nu-n}{2}+1) \Gamma(\frac{1-\nu-n}{2})}\,,
\end{equation}
and $\Gamma(x)$ is the gamma function, \cite{ArfkenBook}.
\\
In this case $x= \cos \theta = \sin \ell$, so the required powers $x^n = \sin^{n} \ell$ can be calculated using the formulas \cite{Gradshteyn2007}
\begin{subequations} \label{eqssin2nlsin2nm1l}
\begin{eqnarray}
\sin^{2n} \ell &=& \frac{1}{2^{2 n-1}} \sum_{k=1}^{n} (-1)^k \binom{2 n}{n-k} \cos 2 k \ell \,,
 \nonumber \\
&+& \frac{1}{2^{2 n}} \binom{2 n}{n} \\
\sin^{2n-1} \ell &=& \frac{1}{2^{2 n-2}} \sum_{k=1}^{n} (-1)^{k+1} \binom{2 n-1}{n-k}  \sin \left(2 k -1\right) \ell \,. \nonumber \\
\end{eqnarray}
\end{subequations}
Substituting Eqs. (\ref{Eq:Pnuxeq0}) and (\ref{eqssin2nlsin2nm1l}) in Eq. (\ref{TayloSerPnu}), we obtain Eq. (\ref{Eq:trigexpPnsinl}) for the trigonometric expansion of $P_\nu(\sin \ell)$.

\end{appendix}

\bibliography{ref}

\end{document}